\documentclass[12pt]{article}

\usepackage{epsfig}
\usepackage{a4p}

\newcommand{\mulc}{\multicolumn}
\newcommand{\ra}{\rightarrow}
\newcommand{\ee}{{\rm e^+e^-}}

\newcommand{\grav}{\tilde {\rm G}}
\newcommand{\chp}{\tilde \chi^+}
\newcommand{\chm}{\tilde \chi^-}
\newcommand{\chpm}{\tilde \chi^\pm}

\newcommand{\nt}{\tilde \chi^0}

\newcommand{\slepton}{\tilde \ell}

\newcommand{\selectron}{\rm\tilde e}
\newcommand{\smu}{\tilde\mu}
\newcommand{\smuon}{\tilde\mu}
\newcommand{\stau}{\tilde\tau}

\newcommand{\Wrvpm}{{\rm W}^{(*)\pm}}
\newcommand{\Wrvm}{{\rm W}^{(*)-}}
\newcommand{\Wrvp}{{\rm W}^{(*)+}}
\newcommand{\qq}{{\rm q \bar q}}
\newcommand{\mumu}{\mu^+\mu^-}
\newcommand{\tautau}{\tau^+\tau^-}

\newcommand{\WW}{{\rm W^+ W^-}}

\newcommand{\Zv}{{\rm Z}^{(*)}}

\newcommand{\dm}{\Delta m}

\begin{document}

\setlength{\parskip}{\medskipamount}


\begin{titlepage}
\begin{center}{\large
    EUROPEAN ORGANIZATION FOR NUCLEAR RESEARCH
    }
\end{center}
 \bigskip
  \begin{flushright}
    CERN-EP-2000-078\\
      19 May 2000\\
      Revised Journal Version 9 January 2001\\
  \end{flushright}
  \bigskip\bigskip\bigskip\bigskip\bigskip
  \begin{center}{\huge\bf\boldmath
      Searches for Prompt Light Gravitino Signatures\\
      in e$^+$e$^-$ Collisions
      at $\sqrt{s} =$~189~GeV
      }
  \end{center}
  \bigskip\bigskip
  \begin{center}{\LARGE
      The OPAL Collaboration
      }
  \end{center}
  \bigskip\bigskip

\begin{abstract}
  Searches for final states expected in 
  models with light gravitinos
  have been performed, including
  experimental topologies
  with multi-leptons
  with missing energy, leptons and photons with missing energy,
  and jets and photons with missing energy.
  No excess over the
  expectations from the Standard Model has been observed.
  Limits are placed on production cross-sections in the
  different experimental topologies.
  Additionally, combining with searches for the anomalous
  production of lepton and photon pairs with missing energy,
  results are interpreted in the context of minimal models
  of gauge mediated SUSY breaking.
  Exclusion limits at the 95\% confidence level on the
  supersymmetric particle masses of
  $m_{\slepton}>$~83~GeV and $m_{\nt_1}>$~85~GeV for $\tan\beta=2$,
  and 
  $m_{\stau}>$~69~GeV, $m_{\selectron,\smu}>$~88~GeV
  and $m_{\nt_1}>$~76~GeV for $\tan\beta=20$,
  are established.
\end{abstract}

 \bigskip
 \bigskip
 \begin{center}
   
 {\large \bf
   (Submitted to Phys. Lett. B.)
   }  
 \end{center}
 
\end{titlepage}


\begin{center}{\Large        The OPAL Collaboration
}\end{center}
\begin{center}{\small
G.\thinspace Abbiendi$^{  2}$,
K.\thinspace Ackerstaff$^{  8}$,
C.\thinspace Ainsley$^{  5}$,
P.F.\thinspace Akesson$^{  3}$,
G.\thinspace Alexander$^{ 22}$,
J.\thinspace Allison$^{ 16}$,
K.J.\thinspace Anderson$^{  9}$,
S.\thinspace Arcelli$^{ 17}$,
S.\thinspace Asai$^{ 23}$,
S.F.\thinspace Ashby$^{  1}$
% ...
}\end{center}\bigskip
\bigskip\small
$^{  1}$School of Physics and Astronomy, University of Birmingham,
Birmingham B15 2TT, UK
\newline
$^{  2}$Dipartimento di Fisica dell' Universit\`a di Bologna and INFN,
I-40126 Bologna, Italy
\newline
$^{  3}$Physikalisches Institut, Universit\"at Bonn,
D-53115 Bonn, Germany
\newline
% ...
$^{  5}$Cavendish Laboratory, Cambridge CB3 0HE, UK
\newline
% ...
$^{  8}$CERN, European Organisation for Nuclear Research,
CH-1211 Geneva 23, Switzerland
\newline
$^{  9}$Enrico Fermi Institute and Department of Physics,
University of Chicago, Chicago IL 60637, USA
\newline
% ...
$^{ 16}$Department of Physics, Schuster Laboratory, The University,
Manchester M13 9PL, UK
\newline
$^{ 17}$Department of Physics, University of Maryland,
College Park, MD 20742, USA
\newline
% ...
$^{ 22}$Department of Physics and Astronomy, Tel Aviv University,
Tel Aviv 69978, Israel
\newline
$^{ 23}$International Centre for Elementary Particle Physics and
Department of Physics, University of Tokyo, Tokyo 113-0033, and
Kobe University, Kobe 657-8501, Japan
\newline
% ...
Budapest, Hungary.


\section{Introduction}

\normalsize

\label{s:intro}

Supersymmetry (SUSY) 
provides a method of solving the ``naturalness'' or ``hierarchy'' problem
by introducing
a set of new particles which cancel the large radiative corrections
to the Higgs mass.
The cancellation is achieved by assuming that, for each
Standard Model (SM)
particle chirality state, there is one additional particle, identical
to its SM partner except that its spin differs by 1/2 unit.
If SUSY were an exact symmetry, the new SUSY particles would
have the same masses as their SM partners.  Since this scenario is
experimentally excluded, SUSY must be a broken symmetry.  It is
typically assumed that SUSY is broken in some ``hidden'' sector
of new particles, and is ``communicated'' (or mediated) to the
``visible'' sector of SM and SUSY particles by one of the known
interactions.  Two scenarios for this mediation have been widely
investigated:  gravity and gauge mediation.
In gauge mediated SUSY breaking (GMSB), the
hidden\footnote{In some GMSB papers this is called the
  ``secluded'' sector, to avoid confusion with the hidden sector
  of gravity mediated SUSY breaking models.}
sector can lie at
energies as low as about $10^4$~GeV.
In most current GMSB theoretical work
\cite{ref:dtw,ref:akm,ref:gr},
it is assumed that this hidden sector is coupled to a messenger
sector, which in turn
couples to the visible sector through normal SM gauge
interactions.  The advantage of GMSB over gravity mediated models
is that flavour changing neutral
currents cannot be induced by SUSY breaking because the normal
gauge interactions are flavour blind.
A feature which distinguishes gravity
from gauge mediated models is the mass of the gravitino,
$\rm \tilde G$.  In gravity mediated models, $\rm \tilde G$ is
usually too heavy to have a significant effect on SUSY phenomenology,
while in GMSB models, the $\rm \tilde G$ is typically
light ($<$~1~GeV) and is the lightest SUSY particle, the LSP.
While $\rm \tilde G$ is a spin 3/2 particle, only its spin 1/2
component (which has ``absorbed'' the goldstino associated with
spontaneous SUSY breaking via the ``superhiggs'' mechanism) interacts
with weak, rather than gravitational, strength interactions, and
contributes to phenomenology.

The next-to-lightest SUSY particle (NLSP) is usually
either the lightest neutralino ($\tilde \chi^0_1$) or the
lightest scalar lepton ($\rm \tilde \ell^\pm_1$), and in a
significant
fraction of the parameter space the NLSP is the lightest scalar
tau lepton ($\rm \tilde \tau_1$).
The coupling of the SUSY particles to $\grav$ is small, and
typically SUSY particles will decay to the NLSP, which then
decays to the gravitino via $\nt_1\ra\gamma\grav$ or
$\slepton\ra\ell\grav$.
If the decay to the gravitino occurs with a small lifetime,
the distinguishing feature 
is events with energetic leptons or photons, plus
significant missing energy due to the missing gravitinos.
OPAL has considered scalar lepton
and lightest neutralino pair creation in these
scenarios in previous publications \cite{ref:acopll189,ref:ggOPAL189}.
This paper reports the first OPAL results on the systematic search for
experimental topologies expected in SUSY models
with a light gravitino, assuming prompt
decays of the NLSP into the $\rm \tilde G$.
It is also possible for the NLSP lifetime to be
significant and it may decay near the interaction point, at an observably
macroscopic distance, or outside the detector.
Many signatures expected with long NLSP lifetimes have been
considered in other OPAL publications
\cite{ref:acopll189,ref:chn189,ref:heavystable183},
and the case of arbitrary lifetime will be considered in a
subsequent publication.
In addition to the new experimental searches
which have sensitivity to general SUSY models
with light gravitinos,
the results reported in this paper are also combined
with those from previous publications
to constrain minimal GMSB models.
Results from searches for GMSB have also been reported by
other collaborations \cite{ref:alephgmsb,ref:delphigmsb}.


\section{OPAL Detector and Event Simulation}

The OPAL detector is described in detail in
Reference~\cite{ref:OPAL-detector}.
The SUSYGEN \cite{ref:susygen} event generator was used to simulate
most of the signal events.  For $\nt_1$ pair creation
in the $\nt_1$ NLSP case, 1000 events
were simulated for each $\nt_1$ mass, using 6~$m_{\nt_1}$ points
from 50~GeV to 94~GeV.  For $\slepton$ pair production in the
$\nt_1$ NLSP case, 1000 events were generated for each
($m_{\slepton}$,$m_{\nt_1}$), using a grid of 48 different points
for each $\slepton$ flavour.  Similar one- or two-dimensional
mass grids were used for $\slepton$ and $\nt_1$ pair production
in the $\slepton$ (and $\stau$) NLSP cases.
For chargino pair production, $\chp_1\chm_1$, and the associated
pair production of the lightest and second lightest neutralino,
$\nt_2\nt_1$, in the $\nt_1$ NLSP case,
the W and Z boson widths can play an important
role and are not fully treated in SUSYGEN.
The DFGT generator \cite{ref:DFGT} 
is used to simulate these signal events.
It includes spin correlations and
allows for a proper treatment of the W boson and the Z boson 
width effects in the chargino and heavy neutralino decays.
Both SUSYGEN and DFGT include initial-state radiation.
The JETSET~7.4 package~\cite{ref:PYTHIA}
is used for the hadronization.
The gravitino mass is set identically to zero in the generation,
since a small mass in the range favoured by the models
has a negligible effect on the detection efficiencies.

The sources of background
include two-photon, lepton-pair, multihadronic, and four-fermion
processes.
The Monte Carlo generators
PHOJET~\cite{ref:PHOJET} (for $Q^2 < 4.5$~GeV$^2$) and 
HERWIG~\cite{ref:HERWIG} (for $Q^2 \geq 4.5$~GeV$^2$)
are used to simulate hadronic events from two-photon
processes.
The Vermaseren~\cite{ref:Vermaseren} program is used to
simulate leptonic two-photon processes  
($\ee\ee$, $\ee \mumu$ and $\ee \tautau$). 
Four-fermion processes 
were simulated using KORALW~\cite{ref:koralw}, and
with the grc4f~\cite{ref:grc4f} generator,
both of
which take into account all interfering four-fermion diagrams.   
The improved simulation of
the transverse momentum of photons from
initial-state radiation makes
the use of KORALW essential for events including photons in the
final states.
Lepton pairs were simulated using
the KORALZ~\cite{ref:KORALZ} generator for
$\tau^+ \tau^- (\gamma)$, $\mumu (\gamma)$ and $\nu\bar\nu\gamma(\gamma)$ events,
and the BHWIDE~\cite{ref:BHWIDE} (when both the electron and
positron scatter at least 12.5$^\circ$ from the beam axis)
and TEEGG~\cite{ref:teegg} (for the remaining phase space) programs
for $\ee \ra \ee (\gamma) $ events.
Multihadronic, $\qq (\gamma)$, events were simulated using
PYTHIA~\cite{ref:PYTHIA}. 

Generated signal and background events were processed
through the full simulation of the OPAL detector~\cite{ref:GOPAL}
and the same event analysis chain was applied to the simulated events
as to the data.  A data set of approximately 182~pb$^{-1}$ at
a luminosity weighted centre of mass energy
of $\sqrt{s}=188.7$~GeV was used for the analysis.


\section{Analysis}
\label{s:analysis}

For all the selections, after the event reconstruction,
double-counting of energy between tracks and calorimeter
clusters is corrected by reducing the calorimeter cluster energy
by the expected energy deposition from aligned
tracks \cite{ref:MT}.
The selections for events with leptons or hadronic jets plus photons and
missing energy described in Section~\ref{ss:ntnlspanal},
as well as four or six leptons plus missing energy in
Section~\ref{ss:slnlspanal}, use the same
lepton
identification
and isolation requirements and preselection as the OPAL Chargino/Neutralino
analysis \cite{ref:chn189}.  The most significant preselection cuts
require that there is no significant energy in the OPAL forward
calorimeters.


\subsection{\boldmath $\nt_1$ NLSP}
\label{ss:ntnlspanal}


\subsubsection{\boldmath $\nt_1\nt_1$ Production with $\nt_1$ NLSP}
\label{sss:n1n1_ntnlspanal}

The search for lightest neutralino pair production followed
by the decays $\nt_1\ra\gamma\grav$ uses the OPAL selection of
events with photon pairs and missing energy \cite{ref:ggOPAL189}.
The analysis selects events with at least two photon candidates
and significant missing energy, along with no other
significant energy in the event.  A total of 24 events are
selected, which is consistent with the expectation of 26.9$\pm$1.2
events from Standard Model $\ee\ra\nu\bar\nu\gamma\gamma(\gamma)$
production and 0.11$\pm$0.04 from all other sources.  The selection
efficiency for $\ee\ra\nu\bar\nu\gamma\gamma(\gamma)$ within the
kinematic acceptance of the analysis is (66.4$\pm$2.9)\%.
One can calculate \cite{ref:gravitinos2} the maximum neutralino mass,
$M_{\nt_1}^{\rm max}$, which is
consistent with the measured three-momenta of the two photons.
A cut on $M_{\nt_1}^{\rm max}$
provides further suppression of the $\nu\bar\nu\gamma\gamma(\gamma)$ 
background while retaining high efficiency for the  signal hypothesis. 
We require that the maximum kinematically allowed mass be greater than
$m_{\nt_1}-5$ GeV,
which retains $(95.5^{+2.0}_{-1.0})$\% relative efficiency for
signal at all values 
of $m_{\nt_1}$ while suppressing much of the remaining
$\nu\bar\nu\gamma\gamma(\gamma)$ background. 
The number of selected events consistent with a given 
value of $m_{\nt_1}$ varies from 14 for $m_{\nt_1}\geq$ 45 GeV
to 3 events at the kinematic limit.
The expected number of SM background events decreases from
$13.67 \pm 0.20$ at $m_{\nt_1}\geq 45$ GeV 
to $1.34 \pm 0.07$ consistent with $m_{\nt_1} \geq 94$ GeV.


\subsubsection{\boldmath $\slepton^+\slepton^-$, $\chp_1\chm_1$ and
                         $\nt_2\nt_1$ Production with $\nt_1$ NLSP}
\label{sss:gX_ntnlspanal}

With a $\nt_1$ NLSP, scalar lepton, charginos and neutralinos
may be observed via\\
$\ee\ra\slepton^+\slepton^-\ra(\ell^+\nt_1)(\ell^-\nt_1)\ra(\ell^+\gamma\grav)(\ell^-\gamma\grav)$,\\
$\ee\ra\chp_1\chm_1\ra(\Wrvp\nt_1)(\Wrvm\nt_1)\ra(\Wrvp\gamma\grav)(\Wrvm\gamma\grav)$
 and\\
$\ee\ra\nt_2\nt_1\ra\Zv\nt_1\nt_1\ra(\Zv\gamma\grav)(\gamma\grav)$.\\
In all cases, the signature is events with two photons plus missing
energy, plus other activity in the detector.  Scalar lepton
production will always lead to a low multiplicity final state,
and only events with at most six tracks are considered
in the analysis.
Chargino pair production and neutralino associated pair production
may lead to either low or high multiplicity final states depending
on the decays of the $\Wrvpm$ and $\Zv$ bosons.  For charginos
and neutralinos, the analysis is therefore divided into two
categories:
\begin{itemize}
  \item[(HM)] High-multiplicity topologies, with
             $\rm N_{\rm ch} - N_{conv} > 4$, where $\rm N_{\rm ch}$
             is the total number of tracks in the event, and
             $\rm N_{conv}$ is the number of tracks originating
             from identified photon conversions,
  \item[(LM)] Low-multiplicity topologies, with
             $\rm N_{\rm ch} - N_{conv} \leq 4$.
\end{itemize}
The background composition depends on the event kinematics, which
are functions of the mass difference between
the produced particles and the lightest neutralino,
$\dm=m-m_{\nt_1}$.
The analysis is therefore separately optimized for different $\dm$
regions, listed in Table~\ref{t:cuts2g}.

The analyses select events with significant missing energy
and two photons.  Cuts are applied on the event
acoplanarity\footnote{After forcing the event into two jets, the acoplanarity
  angle is 180$^\circ$ minus the opening angle between the jets in
  the plane transverse to the beam axis.} ($\phi_{\rm acop}$),
polar angle of the missing momentum ($\cos\theta_{\rm miss}$),
missing transverse momentum scaled by the beam energy
($p_T^{\rm miss}/E_{\rm beam}$) and
visible energy scaled by the centre-of-mass energy ($E_{\rm vis}/\sqrt{s}$).
Additionally, unlike other SUSY searches,
the LSP gravitino is essentially massless,
and the backgrounds can be further reduced while
retaining high efficiency by also imposing a minimum
requirement on $E_{\rm vis}/\sqrt{s}$.
In the HM analyses, there is significant background from
$\ee\ra\WW$ in some of the kinematic regions.  This background
is reduced by removing the most energetic photon
and forcing the event into two jets, and then cutting
on the two-jet mass ($M_{\rm 2jet}$).
In the scalar lepton search, it is also required that there be at least
one identified, isolated lepton in each event.  To maintain
a general search, if two leptons are found they are not required
to be of the same flavour.

Finally, in all channels, it is required that there be at least two
energetic photons in each event.
Photons are identified by selecting unassociated
clusters in the electromagnetic calorimeter,
with the following isolation requirements in a
cone centred on the cluster
(15$^\circ$ half-angle for the highest energy photon
and 10$^\circ$ half-angle for the second photon):
\begin{itemize}
  \item[$\bullet$] scalar momentum sum $<$ 1 GeV;
  \item[$\bullet$] additional electromagnetic calorimeter energy
            sum $<$ 5 GeV;
  \item[$\bullet$] hadronic calorimeter energy $<$ 5 GeV.
\end{itemize}
The energies of the most ($E_{\gamma 1}$) and second most
($E_{\gamma 2}$) energetic photon are also used to select signal events.

In the $\slepton^+\slepton^-$ selections, $\phi_{\rm acop}>5^\circ$
is required, while the cut is tightened to $\phi_{\rm acop}>10^\circ$
for the $\chp_1\chm_1$ and $\nt_2\nt_1$ selections.  In all channels,
$|\cos\theta_{\rm miss}|<0.95$ is required.  
The values of all other analysis cuts are summarized in
Table~\ref{t:cuts2g}, for the different $\Delta m$ regions.

\begin{table}
\centering
\begin{tabular}{|c|c|c|c|c|c|c|} \hline
   \mulc{2}{|c|}{Region} &
  $p_T^{\rm miss}/E_{\rm beam}$ & $E_{\rm vis}/\sqrt{s}$ &
  $M_{\rm 2jet}$ & $E_{\gamma1}$ & $E_{\gamma2}$ \\
   \mulc{2}{|c|}{}      &
                                &                        &
   (GeV)        &  (GeV)        & (GeV)         \\\hline
   \mulc{7}{|c|}{}   \\
   \mulc{1}{|l}{$\slepton^+\slepton^-$} & \mulc{6}{c|}{}\\\hline
   \mulc{2}{|c|}{3 GeV $<\dm<$10 GeV}
      & $>0.10$ & [0.30,0.80] &       &  $>15$  & $>10$ \\\cline{1-4}
   
   \mulc{2}{|c|}{10 GeV$<\dm<m_{\slepton}/2$}
      & $>0.08$ & [0.40,0.85] &  --   &        & \\\cline{1-2}\cline{4-4}\cline{6-7}
   \mulc{2}{|c|}{$m_{\slepton}/2<\dm<m_{\slepton}$}
      &         & [0.45,0.95] &       &  $>10$ & $>5$\\\hline
   \mulc{7}{|c|}{}   \\
   \mulc{1}{|l}{$\chp_1\chm_1$} & \mulc{6}{c|}{} \\\hline
      & 3 GeV $<\dm<$10 GeV
      & $>0.06$ & [0.30,0.80] & $<60$ &  $>20$ & \\\cline{2-3}\cline{5-6}
   HM & 10 GeV$<\dm<m_{\chpm}/2$
     & $>0.05$ &             &       & [15,70]& \\\cline{2-4}\cline{6-6}
   
      & $m_{\chpm}/2<\dm<m_{\chpm}-20$ GeV
      & $>0.04$ & [0.40,0.90] &       & [10,70]& \\\cline{2-4}\cline{6-6}
      & $m_{\chpm}-20~{\rm GeV}<\dm<m_{\chpm}$
      & $>0.03$ & [0.40,0.95] &       &  [5,50]& $>3$ \\\cline{1-4}\cline{6-6}
      & 3 GeV $<\dm<$10 GeV
      &         &             &    -- &  $>20$ & \\\cline{2-2}\cline{6-6}
   LM & 10 GeV$<\dm<m_{\slepton}/2$
      & $>0.05$ & [0.30,0.80] &       & [15,70]& \\\cline{2-2}\cline{6-6}
   
      & $m_{\chpm}/2<\dm<m_{\chpm}-20$ GeV
      &         &             &       & [10,70]& \\\cline{2-2}\cline{6-6}
   
      & $m_{\chpm}-20~{\rm GeV}<\dm<m_{\chpm}$
      &         &             &       &  [5,50]& \\\hline
   \mulc{7}{|c|}{}   \\
   \mulc{1}{|l}{$\nt_2\nt_1$}  &  \mulc{6}{c|}{} \\\hline      
      & 3 GeV $<\dm<$ 10 GeV
      & $>0.05$ & [0.25,0.75] & $<50$ &  $>20$ & \\\cline{2-5}
   
   HM & 10 GeV$<\dm<$ 30 GeV
      & $>0.04$ & [0.30,0.80] & $<60$ &        & \\\cline{2-6}
    
      & 30 GeV $<\dm<$ 80 GeV
      & $>0.035$& [0.40,0.85] &       & [10,70]& \\\cline{2-4}\cline{6-6}
   
      & 80 GeV $<\dm<M_{\nt_2}$
      & $>0.025$& [0.50,0.90] &       &  [5,60]& $>3$\\\cline{1-4}\cline{6-6}
   
      & 3 GeV $<\dm<$ 10 GeV
      & $>0.05$ & [0.25,0.75] &  --   &  $>20$ & \\\cline{2-4}\cline{6-6}
   
   LM &10 GeV$<\dm<$ 30 GeV 
      & $>0.04$ & [0.30,0.80] &       & [20,70]& \\\cline{2-4}\cline{6-6}
   
      & 30 GeV $<\dm<$ 80 GeV
      & $>0.035$& [0.35,0.85] &       & [10,70]& \\\cline{2-4}\cline{6-6}
   
      & 80 GeV $<\dm<M_{\nt_2}$
      & $>0.025$& [0.35,0.90] &       & [10,60]& \\\hline      
\end{tabular}
\caption[]{\sl
   \protect{\parbox[t]{12cm}{
Analysis requirements on quantities used
in the different $\nt_1$ NLSP selections.
 }} 
}
\label{t:cuts2g}
\end{table}

\begin{figure}[b]
\centerline{\hbox{
    \epsfig{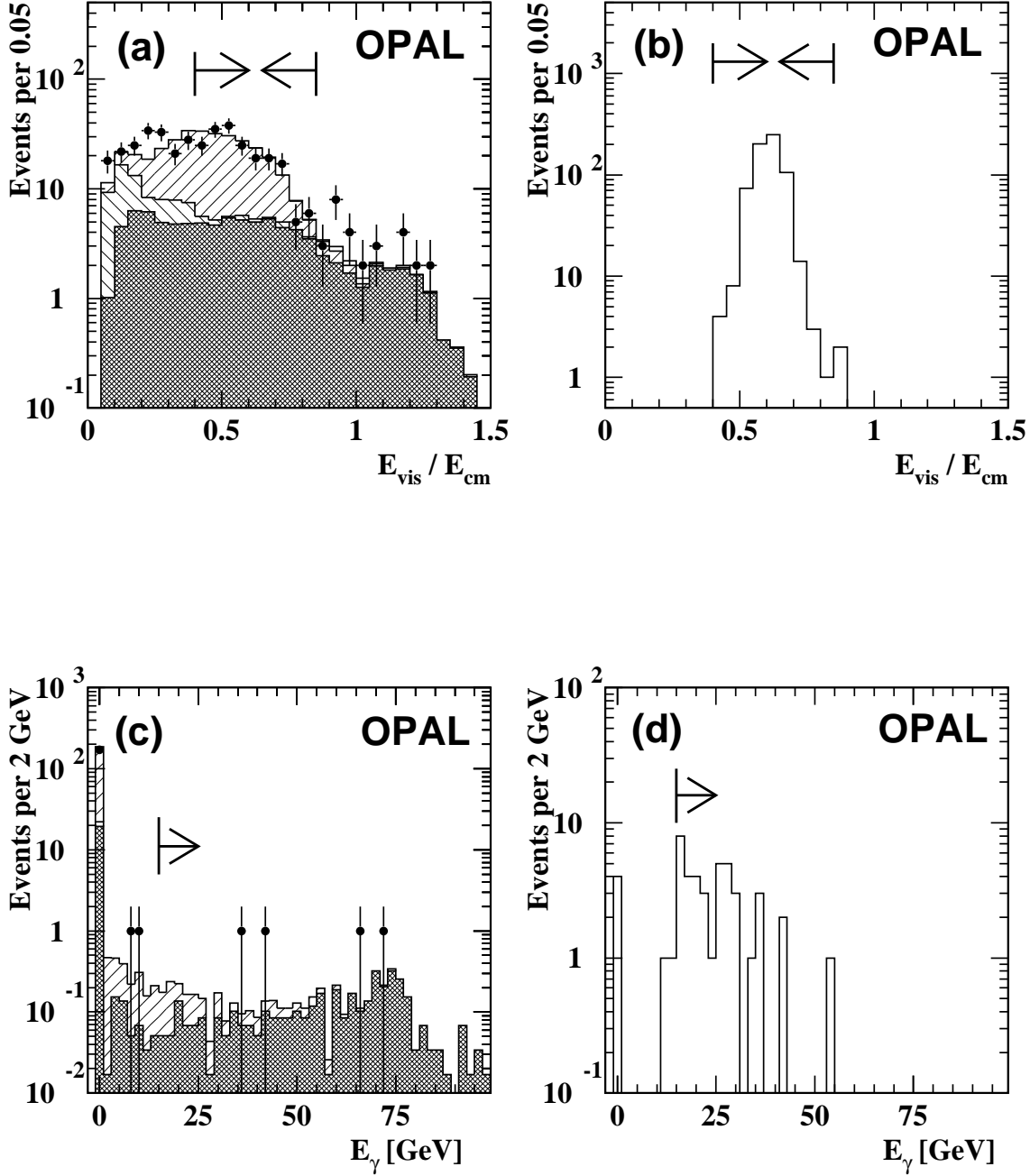}
}}
\caption{
  The distribution of the event visible energy,
  $E_{\rm vis}$,
  and energy of the most energetic photon,
  $E_\gamma$, 
  for events in the $\ell^+\ell^-\gamma\gamma$ analysis
  in Section~\ref{sss:gX_ntnlspanal}.
  In (a) and (c) are shown the data (filled circles with error bars)
  and the prediction from different background processes, normalized
  to the acquired luminosity for the data:  dilepton
  events (double hatched area), two-photon processes (negative slope
  hatching) and four-fermion processes (positive slope hatching).
  In (b) and (d) the prediction for simulated selectron events are shown for
  $m_{\selectron}=$~94~GeV and with $m_{\nt_1}=$~49~GeV.
  The normalization of the signal distribution is arbitrary.
  The arrows indicate the region accepted by the analysis.
}
\label{f:evis_llgg}
\end{figure}

Examples distributions of the event visible energy and the
energy of the most energetic photon are shown in Figure~\ref{f:evis_llgg}.
The selection results are summarized in Table~\ref{t:nev2g},
and the numbers of selected events are consistent with Standard
Model sources.

The detection efficiencies for events from $\slepton^+\slepton^-$
production assuming the decays
$\slepton\ra\ell\nt_1$ and  $\nt_1\ra\gamma\grav$
are typically 30--50\% for scalar electrons and muons,
and 20--40\% for scalar tau leptons.
Summing the high- and low-multiplicity selections,
the detection efficiencies for events from $\chp_1\chm_1$
production assuming the decays
$\chpm_1\ra\Wrvpm\nt_1$ and  $\nt_1\ra\gamma\grav$
range from 20--50\%, depending on the masses of the
chargino and lightest neutralino.
Summing the high- and low-multiplicity selections,
the detection efficiencies for events from $\nt_2\nt_1$
production assuming the decays
$\nt_2\ra\Zv\nt_1$ and $\nt_1\ra\gamma\grav$
are typically 20--50\%, depending on the mass
of the two neutralinos.

\begin{table}
\centering
\begin{tabular}{|c|cc|c|r|} \hline
         Channel         &           Region                       &     &   data    &    total bkg. \\\hline
                         &         3 GeV $<\dm<$10 GeV            &     &     0     &  1.1$\pm$0.2  \\\cline{2-5}
  $\slepton^+\slepton^-$ &    10 GeV$<\dm<m_{\slepton}/2$         &     &     0     &  1.3$\pm$0.2  \\\cline{2-5}
                         & $m_{\slepton}/2<\dm<m_{\slepton}$      &     &     0     &  2.6$\pm$0.4  \\\hline
                         &       3 GeV $<\dm<$10 GeV              & HM  &     0     &  0.6$\pm$0.1  \\
                         &                                        & LM  &     0     &  1.5$\pm$0.4  \\\cline{2-5}
                         &    10 GeV$<\dm<m_{\chpm}/2$            & HM  &     3     &  3.2$\pm$0.8  \\
 $\chp_1\chm_1$          &                                        & LM  &     0     &  2.1$\pm$0.5  \\\cline{2-5}
                         & $m_{\chpm}/2<\dm<m_{\chpm}-20$ GeV     & HM  &     4     &  5.5$\pm$1.7  \\
                         &                                        & LM  &     0     &  1.4$\pm$0.4  \\\cline{2-5}
                         & $m_{\chpm}-20~{\rm GeV}<\dm<m_{\chpm}$ & HM  &     5     &  7.2$\pm$3.4  \\
                         &                                        & LM  &     0     &  0.8$\pm$0.3  \\\hline
                         &  3 GeV $<\dm<$ 10 GeV                  & HM  &     0     &  0.3$\pm$0.05 \\
                         &                                        & LM  &     0     &  1.8$\pm$0.4  \\\cline{2-5}
                         & 10 GeV$<\dm<$ 30 GeV                   & HM  &     0     &  0.7$\pm$0.05 \\
 $\nt_2\nt_1$            &                                        & LM  &     0     &  1.5$\pm$0.4  \\\cline{2-5}
                         & 30 GeV $<\dm<$ 80 GeV                  & HM  &     4     &  4.5$\pm$1.5  \\
                         &                                        & LM  &     0     &  1.7$\pm$0.5  \\\cline{2-5}
                         & 80 GeV $<\dm<$ 180 GeV                 & HM  &     3     &  4.6$\pm$1.6  \\
                         &                                        & LM  &     0     &  1.3$\pm$0.4  \\\hline
\end{tabular}
\caption[]{\sl
   \protect{\parbox[t]{12cm}{
Remaining numbers of events
after all cuts for sleptons, charginos and neutralinos.
There are large correlations among both the selected events
and expected background in the different analyses.
 }} 
}
\label{t:nev2g}
\end{table}


\subsection{\boldmath $\slepton$ NLSP}
\label{ss:slnlspanal}


\subsubsection{\boldmath $\slepton^+\slepton^-$ Production with $\slepton$ NLSP}
\label{sss:slsl_slnlspanal}

The search for lightest scalar lepton pair production followed
by the decays $\slepton^\pm\ra\ell^\pm\grav$ uses the OPAL selection of
events with lepton pairs and missing energy \cite{ref:acopll189}.
The analysis selects events with two lepton candidates
and significant missing energy, along with no other
significant energy in the event.  A total of 301 events were observed
in the data, consistent with the 303.3$\pm$1.9 events expected from
all background sources.
A likelihood selection
using information about the leptons' energies, charges and polar
angles is used to maximize the sensitivity of the analysis to slepton
production for each slepton mass.  An optimized cut for each scalar
lepton mass on the likelihood
is applied, as described in Reference~\cite{ref:acopll183}.
No evidence for anomalous
production of lepton pairs with missing energy is observed.


\subsubsection{\boldmath $\nt_1\nt_1$ Production with $\slepton$ NLSP
  and $\slepton^+\slepton^-$ with $\stau$ NLSP}
\label{sss:multl_slnlspanal}

With a $\slepton$ NLSP, the large neutralino pair production
cross-section may make the 4-lepton plus missing energy signature
$\ee\ra\nt_1\nt_1\ra\slepton\ell\slepton^\prime\ell^\prime\ra(\ell^+\ell^-\grav)(\ell^{\prime+}\ell^{\prime-}\grav)$
the GMSB discovery channel.  Because the scalar tau lepton may be the
lightest slepton, the signature may predominantly include tau leptons.
The selection is sensitive to all 4-lepton plus missing energy final states.
Additionally, with a $\stau$ NLSP,
if the neutralinos are too heavy to be produced and the scalar
tau lepton is significantly lighter than the scalar electron and muon,
then the 6-lepton plus missing energy final state may contribute
via 
$\ee\ra\slepton^+\slepton^-\ra(\ell^+\stau\tau)(\ell^-\stau\tau)\ra(\ell^+\tau^+\tau^-\grav)(\ell^-\tau^+\tau^-\grav)$.

\begin{figure}[b]
\centerline{\hbox{
    \epsfig{file=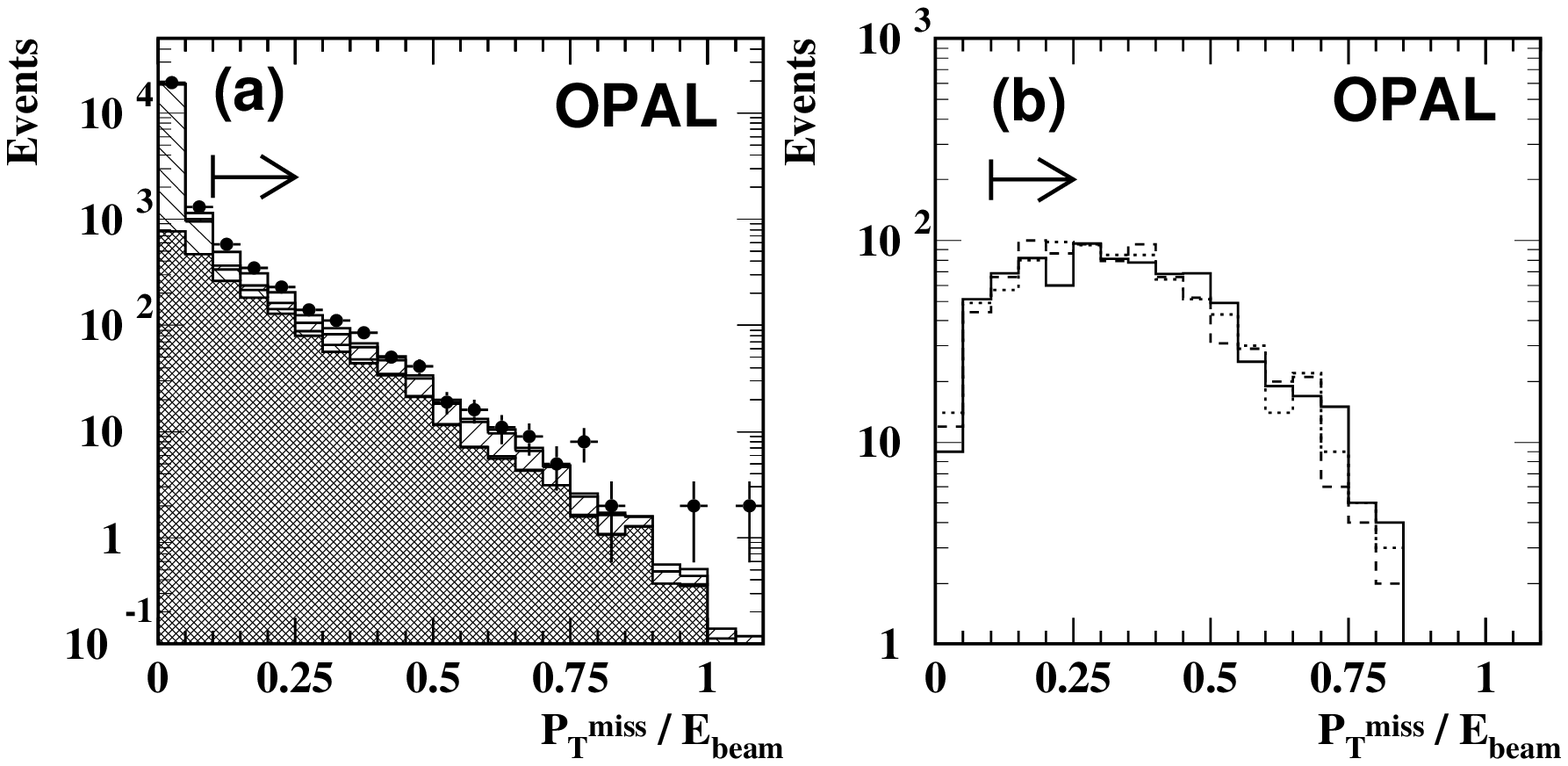,width=\textwidth}
    }}
\centerline{\hbox{
    \epsfig{file=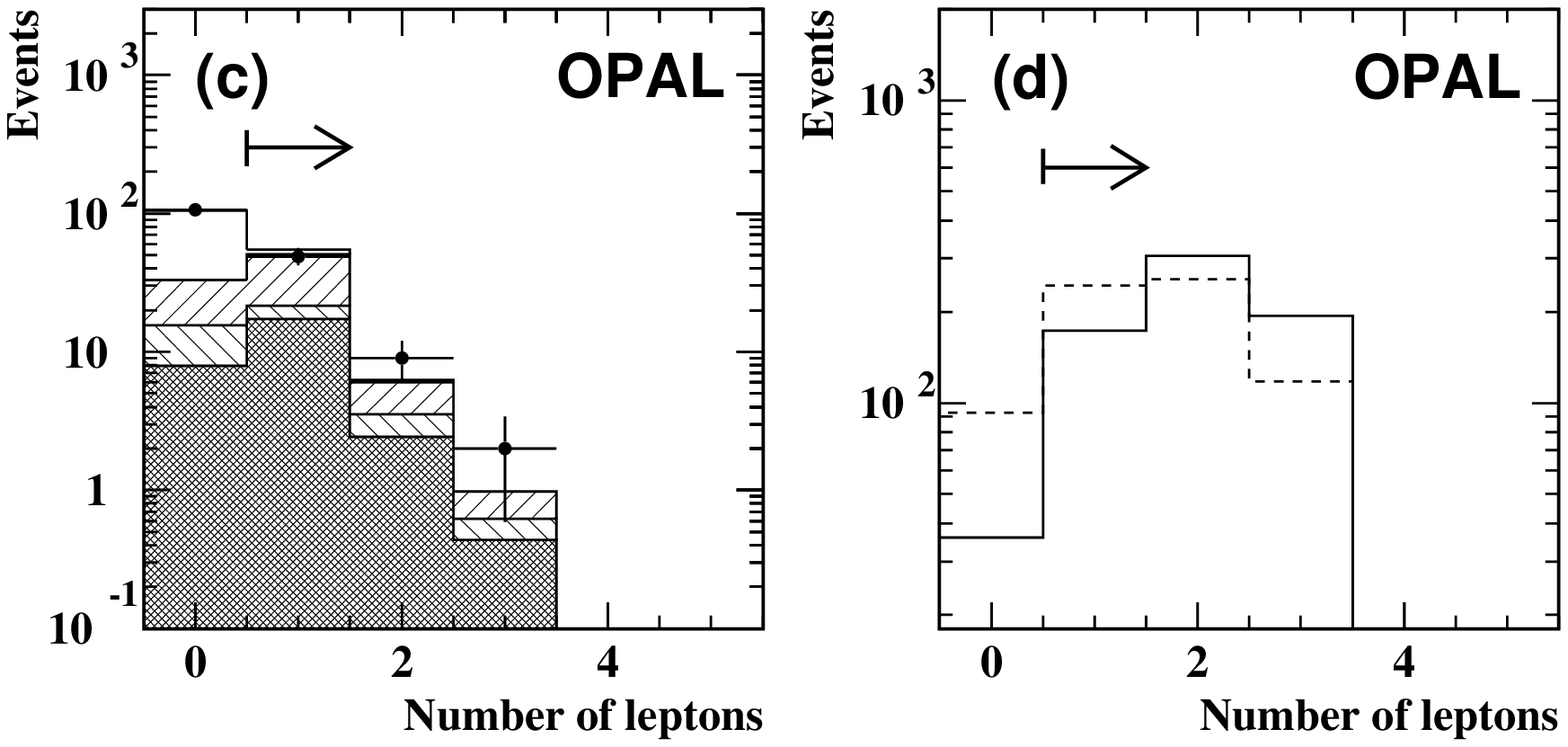,width=\textwidth}
    }}
\caption{
  The event transverse momentum, $p_T^{\rm miss}$,
  for events in the 4 leptons plus missing energy analysis after
  the preselection and requiring between 4 and 10 tracks
  for (a) data and Standard Model backgrounds
  (as for Figure~\ref{f:evis_llgg})
  and (b) predictions from simulated neutralino events for
  $m_{\slepton}=$~50~GeV and with $m_{\nt_1}=$~60~GeV
  (solid line), $m_{\nt_1}=$~90~GeV (dashed line) and with
  $m_{\nt_1}=$~94~GeV (dotted line),
  assuming equal branching fractions for all three lepton generations.
  Also shown are the distributions of the number of identified,
  isolated leptons after the missing energy cuts
  for events in the 6 leptons plus missing energy analysis
  for (c) data and Standard Model backgrounds
  (with an additional contribution from multihadronic
  events shown by the open area)  
  and (d) scalar electron signal Monte Carlo with
  $m_{\selectron}=$~94~GeV and with
  $m_{\stau}=$~85~GeV (solid line) and $m_{\stau}=$~60~GeV
  (dashed line).
  The normalizations of the signal distributions are arbitrary.
  The arrows indicate the region accepted by the analysis.
}
\label{f:multdist}
\end{figure}

The analyses select low multiplicity events by allowing at most
10 tracks.
The events are required to have significant missing energy
by  applying cuts on
$\phi_{\rm acop}$, $\cos\theta_{\rm miss}$,
$p_T^{\rm miss}/E_{\rm beam}$ and $E_{\rm vis}/\sqrt{s}$,
listed in Table~\ref{t:cutsmultl}.
In the 4-lepton
analysis, we require two identified, isolated leptons in the event
and remove Standard Model $\tau^+\tau^-\gamma$ events by vetoing
events with photons with more than half the beam energy.
In the 6-lepton
analysis, only one identified, isolated lepton is required.
This is because these events typically
have only two high energy leptons, both taus, and they often have
nearby tracks from the other decay products and are therefore
not isolated.
If the event is consistent with one lepton plus two hadronic jets
an additional veto on $\rm \ee\ra W^+W^-\ra\qq\ell\nu$ is applied:
events are removed if the invariant mass of the most energetic lepton
and missing momentum is greater than 60~GeV, or if the mass of the
two hadronic jets is greater than 60~GeV.
Examples distributions of the event transverse momentum and the
number of identified, isolated leptons are shown in Figure~\ref{f:multdist}.
The selection results are summarized in Table~\ref{t:nevmultl},
and the numbers of selected events are consistent with Standard
Model sources.

\begin{table}
\centering
\begin{tabular}{|c|c|c|c|c|} \hline
  Channel  & $\phi_{\rm acop}$ & $|\cos\theta_{\rm miss}|$ &
    $p_T^{\rm miss}/E_{\rm beam}$ & $E_{\rm vis}/\sqrt{s}$ \\\hline
$\nt_1\nt_1$
     & $>10^\circ$ & $<0.90$  & $>0.10$  & [0.10,0.90]      \\\cline{1-2}\cline{4-5}
$\slepton^+\slepton^-$
     &    --       &          & $>0.12$  & [0.10,0.80]      \\\hline
\end{tabular}
\caption[]{\sl
   \protect{\parbox[t]{12cm}{
Analysis requirements on quantities used
in the different
$\slepton$ or $\stau$ NLSP selections.
 }} 
}
\label{t:cutsmultl}
\end{table}

\begin{table}
\centering
\begin{tabular}{|c|cc|} \hline
                       Channel                                         & data &    bkg.      \\\hline
   $\nt_1\nt_1\ra(\ell\ell\grav)(\ell^\prime\ell^\prime\grav)$         &   2  &  1.9$\pm$0.2 \\
   $\slepton^+\slepton^-\ra(\ell^+\tau\tau\grav)(\ell^-\tau\tau\grav)$ &   5  &  5.2$\pm$0.4 \\\hline
\end{tabular}
\caption[]{\sl
   \protect{\parbox[t]{12cm}{
The numbers of events remaining after all cuts
in the search for neutralinos and sleptons
with a $\slepton$ NLSP.
 }} 
}
\label{t:nevmultl}
\end{table}

The detection efficiencies for events from $\nt_1\nt_1$ pair production
assuming the decays $\nt_1\ra\slepton\ell\ra\ell\ell\grav$
are fairly uniform for different $\nt_1$ and $\slepton$ masses,
and about 50\% if the neutralino decays
into all three slepton generations with equal branching ratios,
and 35\% if it decays via staus with 100\% branching ratio.
The detection efficiencies for events from selectron and smuon pair
production assuming decays into staus
are also fairly uniform for different sparticle masses,
and are typically about 50\%.


\section{Systematic Errors and Corrections}

Systematic errors on the number of expected signal events arise from
the following sources:  the measurement of the integrated luminosity
(0.5\%); Monte Carlo statistics for the signal samples (1--2\%), and
interpolation errors when determining the efficiencies at arbitrary
masses (typically 5\%);
gaugino field content of the $\chpm$ and $\nt$ which can lead
to different production and decay angular distributions ($< 5$\%);
modelling of the cut variables in the Monte Carlo
simulations (5-10\%).
The cut variable modelling error is determined by shifting each
cut by an amount estimated by comparing data and Monte Carlo
in high statistics samples.

The systematic errors on the expected number of background events
are determined from:  Monte Carlo statistics in the simulated
background events (typically 5\%);
modelling of the cut variables (from 10-20\%, depending on
the analysis and kinematic region).

In the analyses in Sections~\ref{sss:gX_ntnlspanal}
and \ref{sss:multl_slnlspanal},
a common veto on energy
deposition in the forward detectors was applied in the preselection.
The rate of events in which accidental energy depositions in
the forward detectors exceeds the veto thresholds used in the
preselection is estimated from luminosity weighted random beam
crossing events to be 2.9\%.  Since this effect is not included
in the Monte Carlo simulations, the luminosity is reduced
accordingly by this factor when deriving limits using the data.


\section{Results}
\label{s:results}

No significant excesses are observed in any channels,
so limits are derived using the search results.
Limits are derived using the method from Reference~\cite{ref:pdg96bg},
including the effects of systematic errors on the signal
detection efficiencies and background expectation using the
method from Reference~\cite{ref:cousinshighland}.


\subsection{Model Independent Interpretations}
\label{ss:miresults}

For a $\nt_1$ NLSP, assuming
the prompt decay $\nt_1\ra\gamma\grav$,
production cross-section limit contours are calculated
in the mass plane of the particle produced {\it vs.} the
mass of the $\nt_1$.  In this scenario,
limits on the production
cross-sections for the processes
$\ee\ra\smuon^+\smuon^-$,
$\ee\ra\stau^+\stau^-$,
$\ee\ra\chp_1\chm_1$
and $\ee\ra\nt_2\nt_1$ are shown in Figure~\ref{f:xlim_2p}.
Typically, production cross-sections
in excess of about 0.03 -- 0.1~pb are
excluded at the 95\% confidence level.

\begin{figure}[b]
\centerline{\hbox{
    \epsfig{file=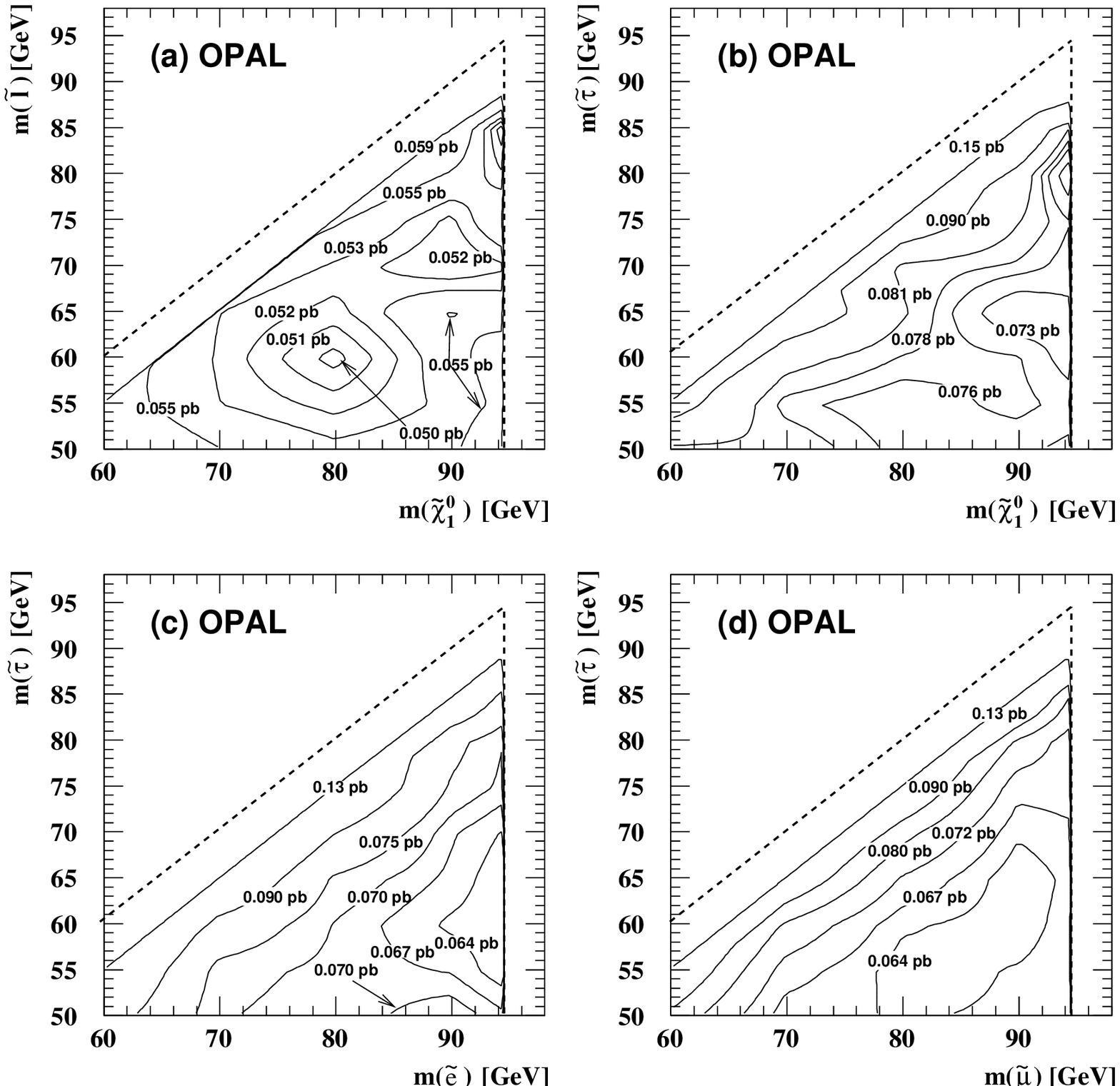,width=\textwidth}
}}
\caption{
  Limits at the 95\% confidence level for the production
  cross-section of $\ee\ra$ (a) $\nt_1\nt_1$ assuming the decays
  $\nt_1\ra\slepton\ell$ with equal branching ratios
  to all three generations, (b) 
  $\nt_1\nt_1$ assuming the decays
  $\nt_1\ra\stau\tau$,
  (c) $\selectron^+\selectron^-$ assuming
  the decays $\selectron\ra\stau\tau{\rm e}$ and
  (d) $\smuon^+\smuon^-$ assuming
  the decays $\smuon\ra\stau\tau\mu$.
  The limits assume that these decays are followed by
  the prompt decays $\slepton\ra\ell\grav$.
  The dashed lines indicate the kinematic limit.
}
\label{f:xlim_multl}
\end{figure}

For a $\slepton$ NLSP, assuming
the prompt decay $\slepton\ra\ell\grav$,
production cross-section limit contours can be calculated
in the mass plane of the particle produced {\it vs.} the
mass of the $\slepton$, shown in Figure~\ref{f:xlim_multl}.
Typically, cross-sections for $\ee\ra\nt_1\nt_1$
larger than 0.05 -- 0.06~pb are excluded at the
95\% confidence level for the degenerate slepton
case, while cross-section larger than
0.07 -- 0.15~pb are excluded for the stau NLSP case.
For scalar electron or muon pair production
with a scalar tau NLSP, the
cross-section limits are typically 0.06 -- 0.13~pb.


\subsection{GMSB Model Dependent Interpretations}
\label{ss:mdresults}

While there is no single GMSB model,
there are typically \cite{ref:dtw,ref:akm,ref:gr}
six new parameters in addition to
those of the SM:
\begin{equation}
  F, ~ \Lambda, ~ M, ~ N, ~ \tan{\beta} ~ {\rm and} ~ {\rm sign}(\mu).
  \label{e:params}
\end{equation}
The intrinsic SUSY breaking scale is $\sqrt{F}$, which also
determines the $\grav$ mass according to
$m_{\grav} \simeq 2.5 \times F / (100~{\rm TeV})^2$~eV.
Since $\sqrt{F}$ affects primarily the lifetime of the NLSP
we do not vary it for this paper, but simply assume that
this lifetime is short enough to have no effect on our
detection efficiencies.
The parameter $\Lambda$ sets
the overall mass scale for SUSY particles,
$M$ is the mass of the messenger
particles, $N$ is the number of
sets\footnote{
  $N$ is technically the Dynkin index of the gauge representation
  of the messenger fields.
  To preserve gauge coupling unification, the messengers are
  assumed to form a GUT representation.  In the simplest
  form, each of the $N$ messenger particle sets has
  the quantum numbers of an
  \mbox{\boldmath $5 + \bar 5$ of $SU(5)$}.  The maximum number
  of messengers can be bounded by requiring the
  gauge interactions remain perturbative
  up the GUT scale,
  although this bound depends on $M$.  For $M=$~100~TeV,
  $N \leq 5$, while for $M=10^{10}$~TeV, $N \leq 10$.
  }
 of messenger particles, and
$\tan{\beta}$ is the usual ratio of the Higgs vacuum
expectation values.
The final parameter is just the sign of the Higgs
sector mixing parameter, $\mu$, which introduces a
two-fold ambiguity (the magnitude of $\mu$ is calculable
from the other parameters in the minimal model by imposing
radiative electroweak symmetry breaking).
The messenger scale gaugino masses can be calculated
using the relation
\begin{equation}
  M_i = N ~ \frac{\alpha_i}{4 \pi} ~ \Lambda 
       ~ g(\Lambda/M),
  \label{e:mgaugino}
\end{equation}
where the index $i$ refers to the $U(1)$, $SU(2)$ or
$SU(3)$ gauge group, and the
$\alpha_i$ are the SM gauge couplings.
The function $g(\Lambda/M)$ is always slightly greater than 1,
but its effect is only significant when $\Lambda\approx M$.
It is also possible to calculate all of the
messenger scale
scalar SUSY particle masses using $\Lambda$ and $N$ via
\begin{equation}
  m^2 = 2 ~ N ~ \Lambda^2 ~ f(\Lambda/M) ~ \sum_{i=1}^{3} k_i
         \left(\frac{\alpha_i}{4 \pi}\right)^2,
  \label{e:msfermion}
\end{equation}
where $k_i$ are multiplicative factors determined by
the particle's SM charge, hypercharge and colour charge.
The function $f(\Lambda/M)$ is usually near 1, except
when $\Lambda\approx M$.
The electroweak scale particle masses may be calculated
from the messenger scale masses using the renormalization
group equations.

We will work in a theoretical framework based on
Reference~\cite{ref:dtw}, extending it by including
a full mass treatment for all three generations of sparticles.
The theoretical calculations are embedded in the
SUSYGEN~\cite{ref:susygen} generator.
The complete interpretation framework is
described in Reference~\cite{ref:duchovnietal}.
The SUSY breaking scale (or equivalently gravitino mass) is
not considered explicitly in this section, although it is
assumed that $M_{\grav}<1$~GeV for the selection efficiencies
to remain valid.  The values of the parameters considered in our
scan are shown in Table~\ref{t:params}.

\begin{table}
\centering
\begin{tabular}{|l|cc|} \hline
     Parameter     &  Lower Value        & Upper Value \\\hline
    $\tan\beta$    &      2              &    50       \\
    $\Lambda$      &    5~TeV            &  200~TeV    \\
    $M$            & 1.01$\times\Lambda$ &  $10^6$ TeV \\
    $N$            &     1               &    4        \\
    ${\rm sign}(\mu)$ & -1               &   +1        \\\hline
\end{tabular}
\caption[]{\sl
   \protect{\parbox[t]{12cm}{
    Parameter ranges considered in GMSB scans.
 }} 
}
\label{t:params}
\end{table}

\begin{figure}[b]
   \centerline{\hbox{
     \epsfig{file=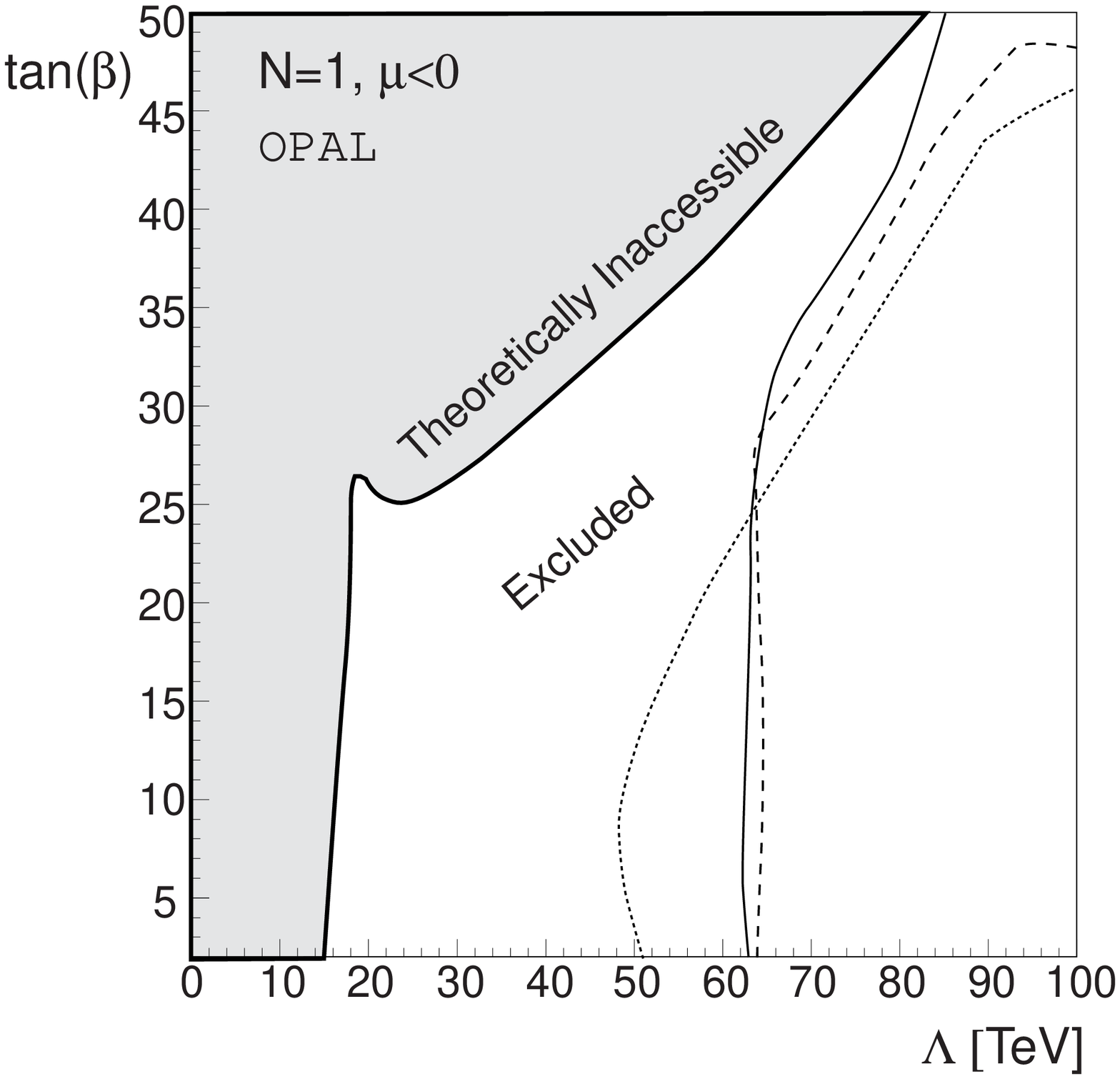,width=0.5\textwidth}
     \epsfig{file=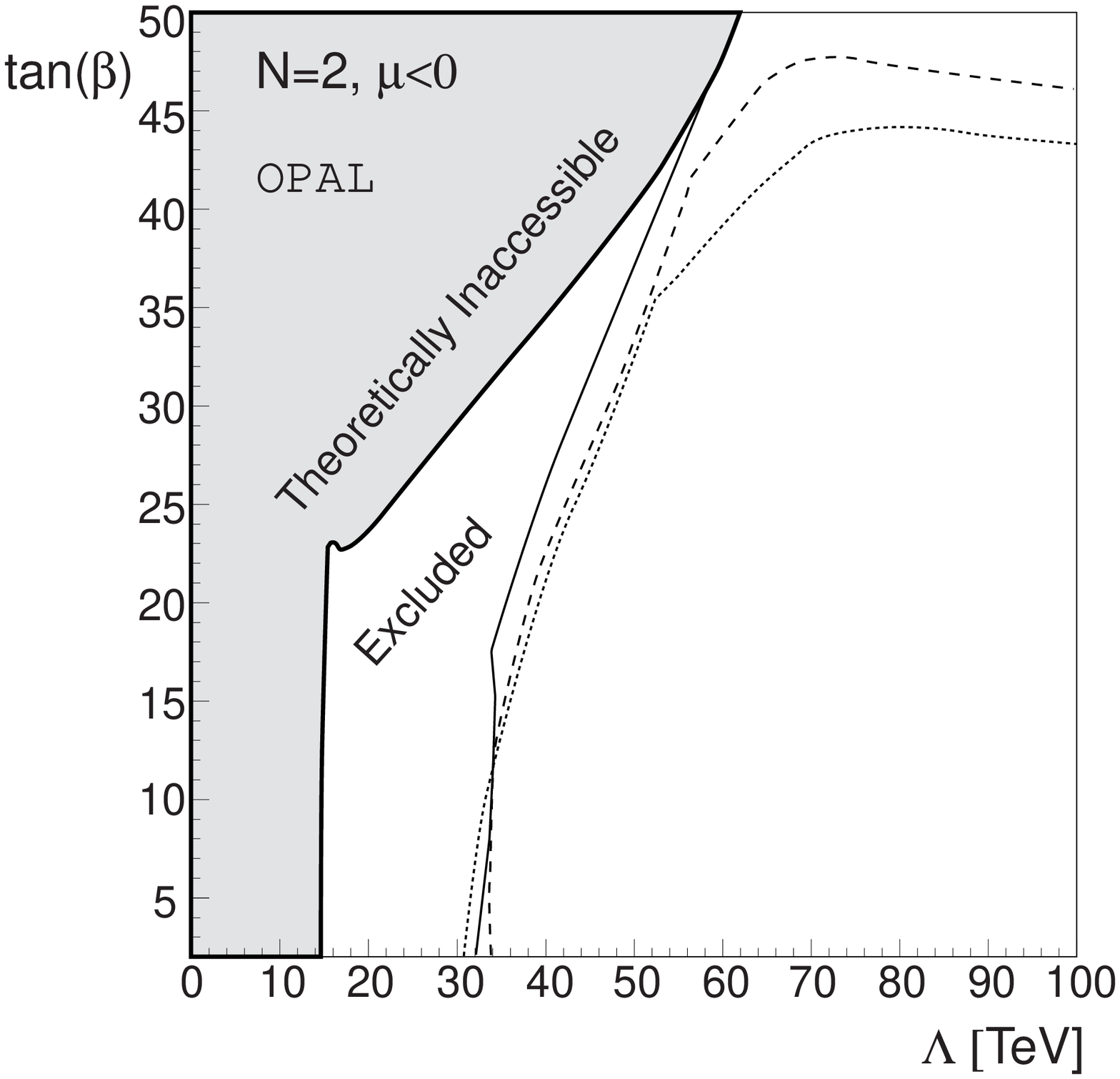,width=0.5\textwidth}
  }}
   \centerline{\hbox{
     \epsfig{file=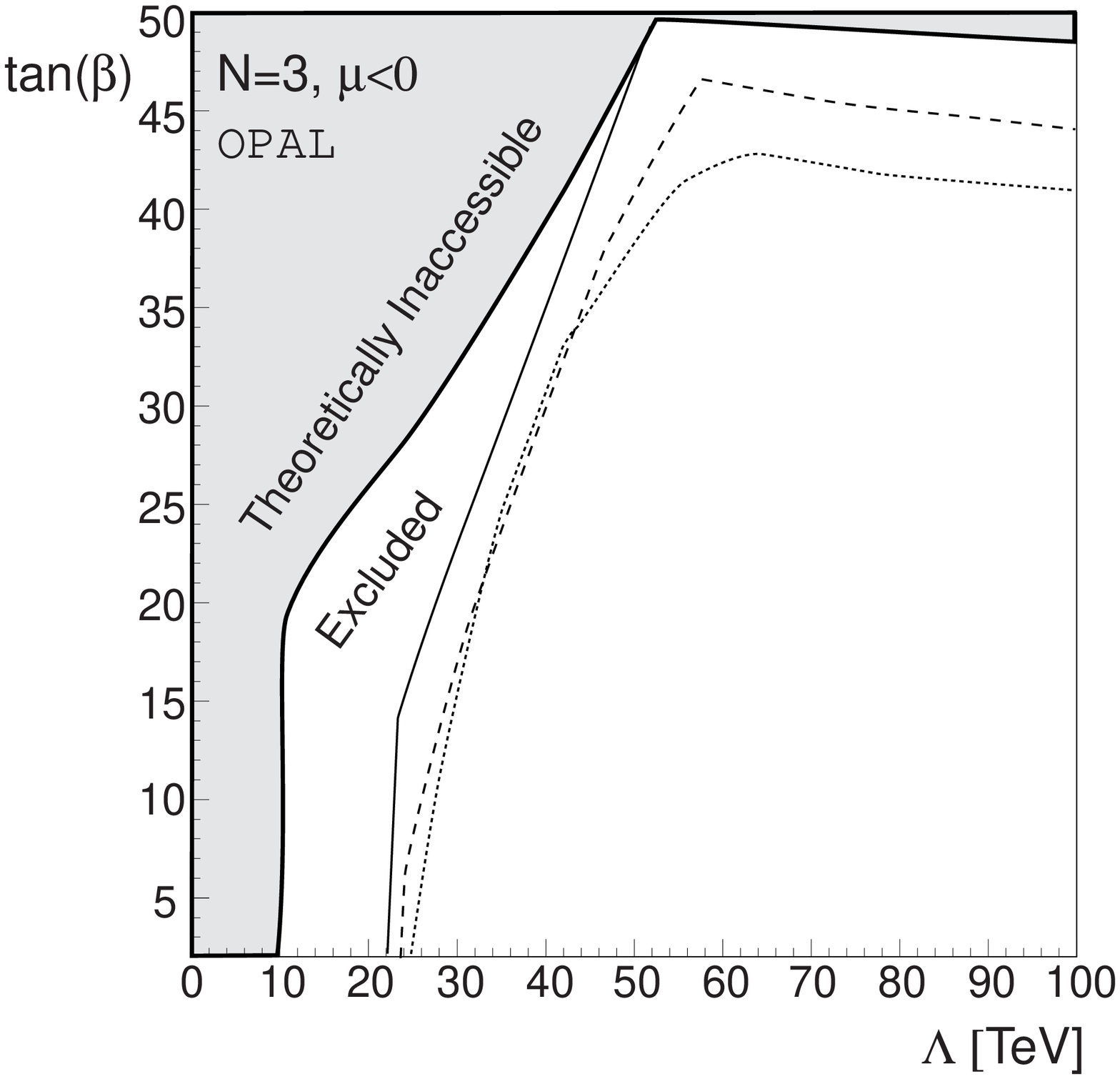,width=0.5\textwidth}
     \epsfig{file=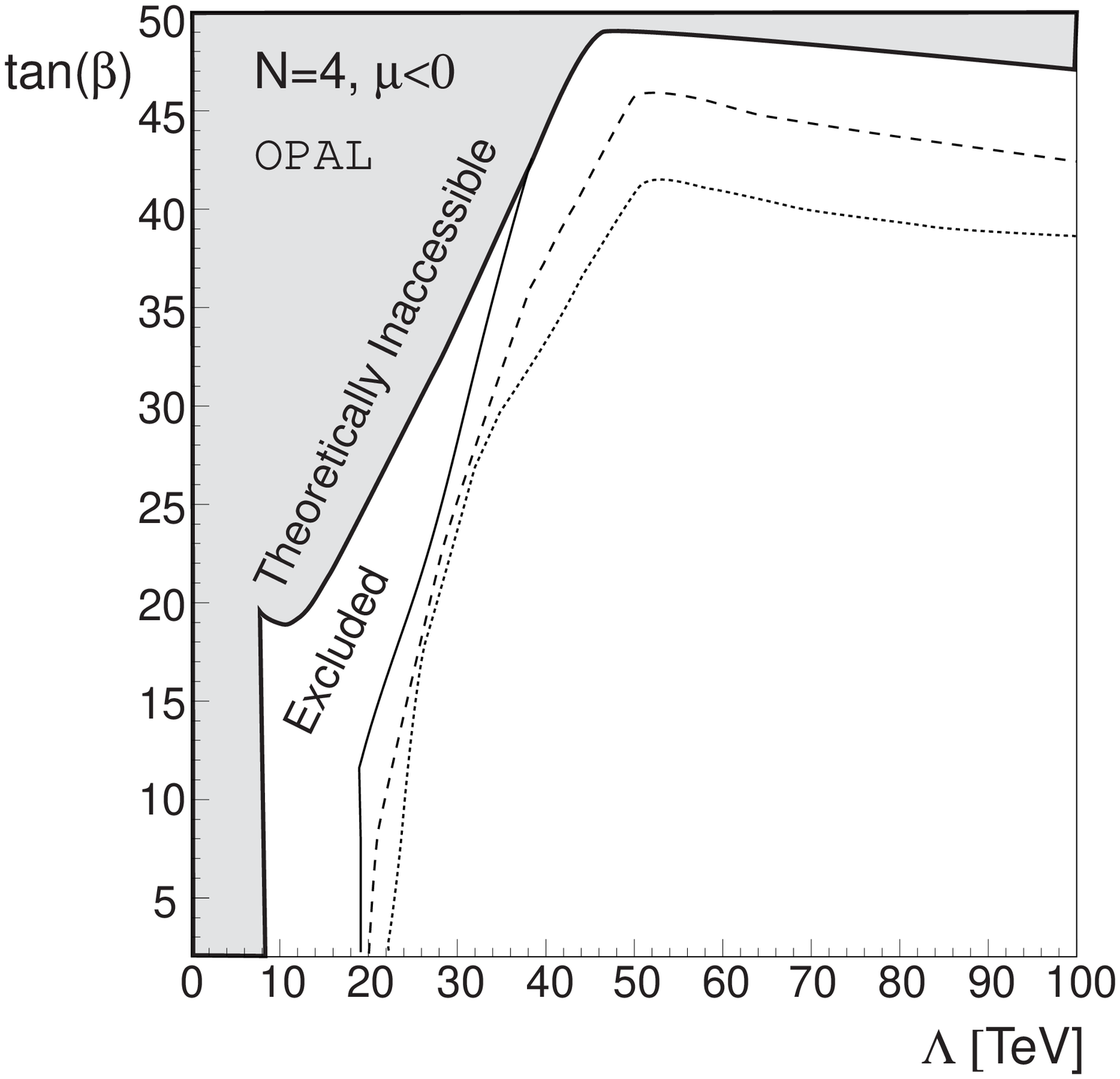,width=0.5\textwidth}
  }}
\caption{
  Excluded regions at the 95\% C.L. in the $\Lambda$ {\it vs.}
  $\tan\beta$ plane for $N=$~1, 2, 3 and 4,
  and for $\mu<0$.
  The areas above and to the left of the solid line are excluded for
  $M=10^{6}$~TeV, the dashed line for $M=$~250~TeV, and
  the dotted line for $M=1.01\times\Lambda$.
  The shaded regions are theoretically
  inaccessible for $M=10^{6}$~TeV (the inaccessible region
  is larger for smaller values of $M$).  The exclusions
  for $\mu>0$ are somewhat stronger.
}
\label{f:Lamvtanb}
\end{figure}

\begin{figure}[b]
   \centerline{\hbox{
     \epsfig{file=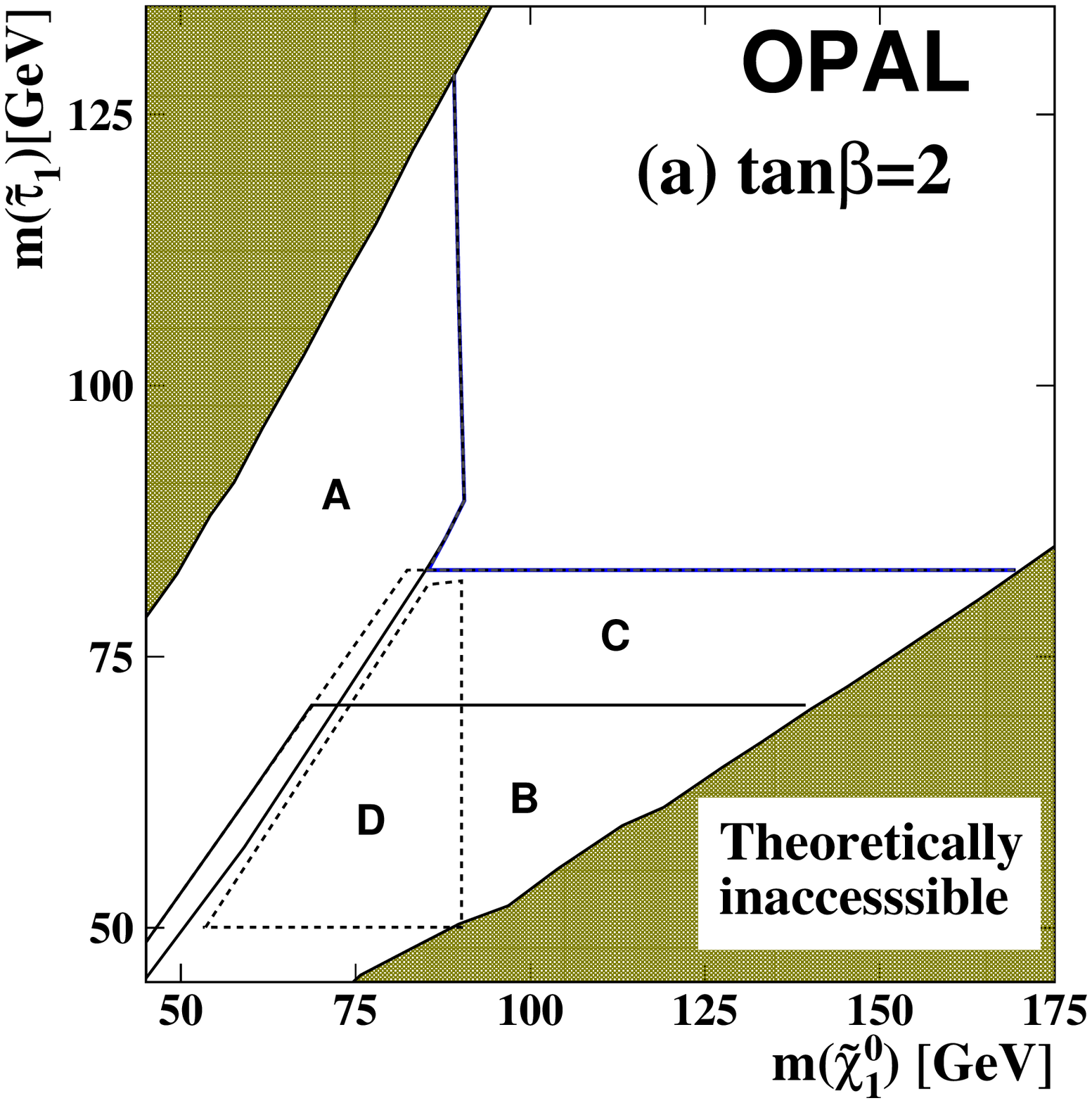,width=0.5\textwidth}
     \epsfig{file=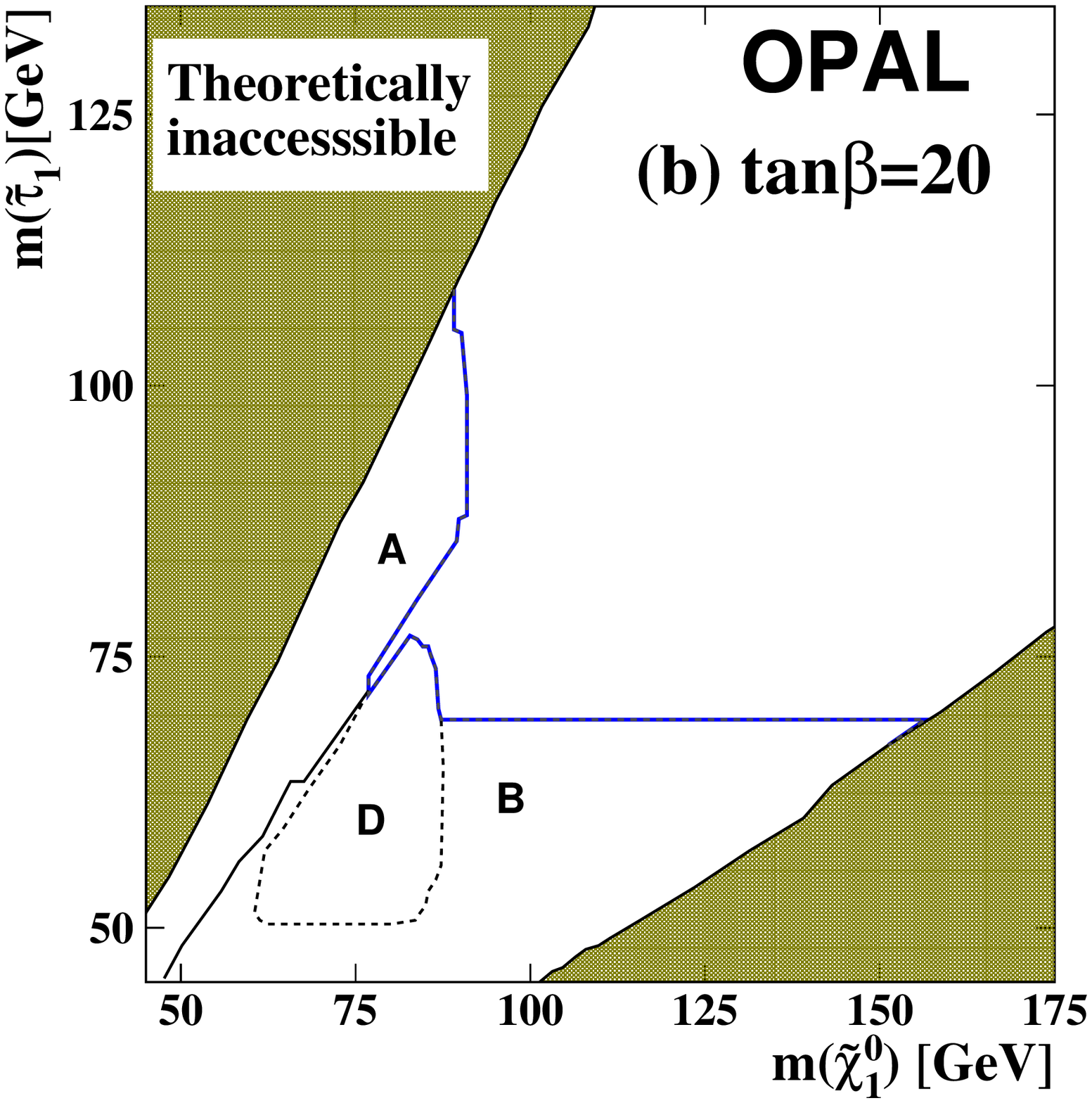,width=0.5\textwidth}
  }}
\caption{
  Excluded regions at the 95\% C.L. in the
  $m_{\slepton}$ {\it vs.} $m_{\nt_1}$ plane for
  (a) $\tan\beta=$~2 (approximately degenerate slepton case)
  and
  (b) $\tan\beta=$~20 (lighter stau case).
  Also shown are the regions exclusively excluded by
      (A) $\nt_1\nt_1\ra\gamma\grav\gamma\grav$,
      (B) $\stau^+\stau^-\ra\tau^+\grav\tau^-\grav$,
      (C) $\smu^+\smu^-\ra\mu^+\grav\mu^-\grav$ and
      (D) $\nt_1\nt_1\ra$~4~lepton final states.
  The other search channels do not contribute significantly
  to the exclusion regions in the minimal model.
  The shaded regions are theoretically inaccessible.
}
\label{f:mslvmchi}
\end{figure}

In Figure~\ref{f:Lamvtanb}, 95\% C.L. exclusion limits in
the $\Lambda$ {\it vs.} $\tan\beta$ plane are shown for
different values of $N$.  The $\nt_1$ NLSP signatures tend
to be dominant for small $N$ and low $\tan\beta$,
while the $\slepton$ NLSP
signatures are more important for either larger $N$ or larger $\tan\beta$.
Absolute 95\% C.L. lower limits on $\Lambda$ of
48, 31, 22 and 19~TeV are established for $N=$~1, 2, 3 and 4, respectively.
In
Figure~\ref{f:mslvmchi}(a), the excluded region in the $M_{\slepton}$
{\it vs.} $M_{\nt_1}$ plane is shown
for $\tan\beta=2$,
corresponding to the case
where the three sleptons are degenerate in mass.
The dominant exclusion channels in the minimal model are
$\nt_1\nt_1\ra\gamma\grav\gamma\grav$
and $\slepton^+\slepton^-\ra\ell^+\grav\ell^-\grav$.
In Figure~\ref{f:mslvmchi}(b), the excluded region in the $M_{\stau}$
{\it vs.} $M_{\nt_1}$ plane is shown
for $\tan\beta=20$, 
corresponding to the case
where the $\stau$ is significantly lighter than the other sleptons.
The $\nt_1\nt_1\ra\gamma\grav\gamma\grav$ channel remains
powerful in this case, but the light  $\stau$ means that
only the $\stau^+\stau^-\ra\tau^+\grav\tau^-\grav$ channel
contributes to the significance from lepton pair final states.

Finally, 95\% C.L. limits can be derived on the NLSP mass of
$M_{\slepton}>$~83~GeV and $M_{\nt_1}>$~85~GeV for $\tan\beta=2$,
and 
$M_{\stau}>$~69~GeV, $M_{\selectron,\smu}>$~88~GeV
and $M_{\nt_1}>$~76~GeV for $\tan\beta=20$.


\section{Conclusion}
\label{s:conclusion}

We have searched for signatures expected in models
with gauge mediated SUSY breaking at a centre-of-mass
energy of $\sqrt{s}=$~189~GeV with the OPAL detector
at LEP.  No evidence in any search channel over the
expectations from the Standard Model was observed.
Limits are placed on the production cross-sections for
a number of processes for the prompt decays of
the next-to-lightest SUSY particle to a gravitino.
The results are used to constrain minimal
models of gauge mediated supersymmetry breaking.


\section{Acknowledgements}
\label{s:acknowledgements}

We particularly wish to thank the SL Division for the efficient operation
of the LEP accelerator at all energies
 and for their continuing close cooperation with
our experimental group.  We thank our colleagues from CEA, DAPNIA/SPP,
CE-Saclay for their efforts over the years on the time-of-flight and trigger
systems which we continue to use.  In addition to the support staff at our own
institutions we are pleased to acknowledge the  \\
Department of Energy, USA, \\
National Science Foundation, USA, \\
Particle Physics and Astronomy Research Council, UK, \\
Natural Sciences and Engineering Research Council, Canada, \\
Israel Science Foundation, administered by the Israel
Academy of Science and Humanities, \\
Minerva Gesellschaft, \\
Benoziyo Center for High Energy Physics,\\
Japanese Ministry of Education, Science and Culture (the
Monbusho) and a grant under the Monbusho International
Science Research Program,\\
Japanese Society for the Promotion of Science (JSPS),\\
German Israeli Bi-national Science Foundation (GIF), \\
Bundesministerium f\"ur Bildung und Forschung, Germany, \\
National Research Council of Canada, \\
Research Corporation, USA,\\
Hungarian Foundation for Scientific Research, OTKA T-029328, 
T023793 and OTKA F-023259.\\



\newpage
\clearpage

\clearpage

\begin{figure}[b]
\centerline{\hbox{
    \epsfig{file=pr309_03.eps,width=\textwidth
  ,bbllx=50pt,bblly=180pt,bburx=570pt,bbury=740pt
  }
}}
\caption{
  Limits at the 95\% confidence level for the production
  cross-section of $\ee\ra$ (a) $\smuon^+\smuon^-$,
  (b) $\stau^+\stau^-$, (c) $\chp_1\chm_1$ and (d) $\nt_2\nt_1$,
  assuming decays via $\nt_1$ followed
  by the prompt decays $\nt_1\ra\gamma\grav$.
  The limits on $\selectron^+\selectron^-$ are essentially
  identical to those for $\smuon^+\smuon^-$.
  The dashed lines indicate the kinematic limit.
}
\label{f:xlim_2p}
\end{figure}
\clearpage

\clearpage

\clearpage


\end{document}